\documentclass[conference]{IEEEtran}
\IEEEoverridecommandlockouts
\usepackage{cite}
\usepackage{graphicx}  
\usepackage{amsmath,amssymb,amsfonts}
\usepackage{algorithmic}
\usepackage{algorithm}
\usepackage{graphicx}
\usepackage{textcomp}
\usepackage{xcolor}

\def\BibTeX{{\rm B\kern-.05em{\sc i\kern-.025em b}\kern-.08em
    T\kern-.1667em\lower.7ex\hbox{E}\kern-.125emX}}
\begin{document}

\title{\huge Efficient Calibration for RRAM-based In-Memory Computing using DoRA
%
}

\author{
Weirong Dong$^{*}$, Kai Zhou$^{*}$, Zhen Kong$^{*}$, Quan Cheng$^{\dagger}$, Junkai Huang$^{*}$,\\ Zhengke Yang$^{*}$, Masanori Hashimoto$^{\dagger}$, Longyang Lin$^{*}$ \\
\IEEEauthorblockA{
    $^*$School of Microelectronics, Southern University of Science and Technology, Shenzhen, China \\
    $^{\dagger}$Department of Informatics, Kyoto University, Kyoto, Japan \\
    linly@sustech.edu.cn
}
}

\maketitle

\begin{abstract}
Resistive In-Memory Computing (RIMC) offers ultra-efficient computation for edge AI but faces accuracy degradation due to RRAM conductance drift over time. Traditional retraining methods are limited by RRAM’s high energy consumption, write latency, and endurance constraints. We propose a DoRA-based calibration framework that restores accuracy by compensating influential weights with minimal calibration parameters stored in SRAM, leaving RRAM weights untouched. This eliminates in-field RRAM writes, ensuring energy-efficient, fast, and reliable calibration. Experiments on RIMC-based ResNet50 (ImageNet-1K) demonstrate 69.53\% accuracy restoration using just 10 calibration samples while updating only 2.34\% of parameters.

\end{abstract}


\section{Introduction}
The success of Deep Neural Networks (DNNs) in a variety of domains has driven the demand for computational efficiency, particularly in edge AI applications where resource constraints, such as limited memory and power, are prevalent. Traditional digital processors, including CPUs and GPUs, are hindered by the ``von Neumann bottleneck,'' where data must be frequently transferred between memory and processing units, leading to latency and energy inefficiencies. In contrast, Resistive Random-Access Memory (RRAM) based In-Memory Computing (RIMC) \cite{RIMC1}\cite{RIMC2}\cite{RIMC3}offers a promising solution by storing and processing data directly within memory cells, thus bypassing the bottleneck and offering significant gains in computational speed and energy efficiency. RIMC leverage the physical properties of resistive devices to encode weight values through conductance, offering an attractive alternative for power-constrained applications, such as those in edge AI and IoT devices.

\begin{figure}[t]
    \centering
    \includegraphics[width=0.5\textwidth, keepaspectratio]{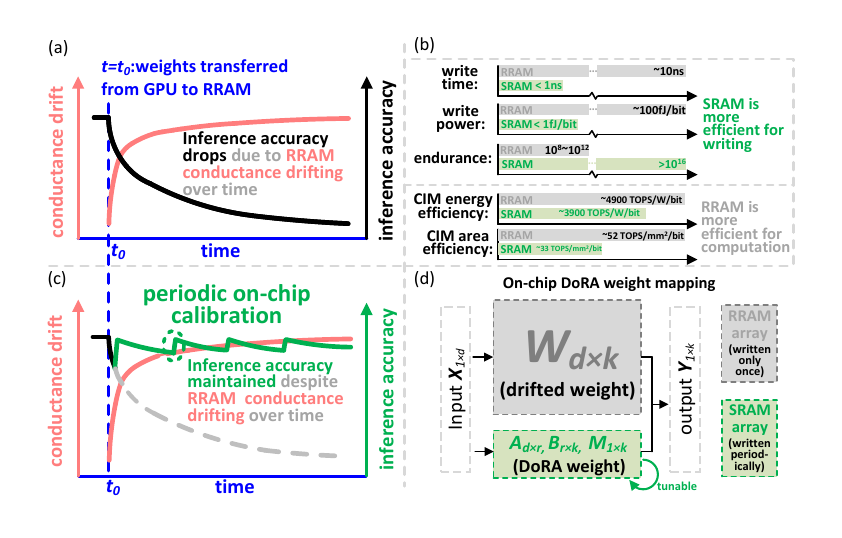}
    \caption{(a) Impact of RRAM conductance relaxation on conductance values and inference accuracy in RIMC-based systems; (b) Comparison of write operations and in-memory computing performance between RRAM and SRAM; (c) Periodic calibration process for accuracy restoration in RIMC systems; (d) Proposed RRAM-friendly calibration approach using the DoRA weight mapping strategy, eliminating the need for RRAM writes. 
    }
    \label{fig1}
\end{figure}

Despite these advantages, the deployment of RRAM for DNNs faces several critical challenges, particularly the conductance relaxation effect \cite{relaxation1}. Over time, the conductance values programmed into RRAM drift due to internal mechanisms such as charge redistribution and interface degradation  \cite{relaxation2}. This relaxation effect causes discrepancies between the actual conductance values  and the target values, directly degrading the accuracy of RRAM-based DNNs. As this drift accumulates, the performance of the neural network deteriorates further, as shown in Fig.~\ref{fig1}(a).

In addition to relaxation, write-related challenges also hinder RRAM deployment as shown in Fig.~\ref{fig1}(b). Programming RRAM conductance values is inherently non-ideal due to device variations and limitations in precision. Achieving the desired target conductance often requires iterative write-and-verify operations \cite{write_and_verify}, which are both time-consuming and energy-intensive. Excessive write operations also exacerbate the endurance limitations of RRAM devices which shorten the device's lifespan \cite{endurance}. Hence, frequent updates in RRAM is impractical for retraining or calibration.

These challenges—programming errors, write endurance limitations, and the conductance relaxation effect—pose significant barriers to the practical deployment of RRAM in DNNs. Existing solutions address these issues partially but fall short of a comprehensive approach. For instance, variation-aware fine-tuning \cite{variationaware} integrates RRAM device variations into GPU-based DNN training to reduce programming errors. However, its variation tolerance is limited, and it cannot effectively counteract conductance drift. RRAM-based training \cite{onchip}\cite{onchip2} addresses both programming errors and conductance drift by calibrating weights directly on RRAM periodically, but it requires numerous write operations, which consume significant time and energy, further exacerbating RRAM's endurance limitations. To date, no approach has simultaneously and efficiently tackled all these challenges.

We propose a lightweight calibration scheme that compensates the drift of weights and consequent accuracy degradation with a limited number of calibration parameters stored in SRAM, where the RRAM arrays are unchanged.
The key contributions of this paper are as follows:

\begin{itemize}
\item We propose a novel calibration method for RIMC systems using Weight-Decomposed Low-Rank Adaptation (DoRA)\cite{DoRA}, shown in Fig.~\ref{fig1}(c-d). This method can restore the inference accuracy by updating only a small subset of weights stored in SRAM, rather than modifying the weights in RRAM. By doing so, we significantly reduce the number of trainable parameters by 95.32\%, and eliminate the need for repetitive RRAM write-and-verify operations, improving calibration efficiency.

\item Inspired by feature-based knowledge distillation (KD) \cite{distillation}\cite{feature}, we propose a feature-based calibration approach that guides the calibration of RRAM-based DNNs layer by layer, using DNNs trained on GPUs. This method minimizes the feature gap (i.e., the output difference of each layer) without relying on cross-layer backpropagation. Remarkably, it requires only 10 calibration samples to achieve high accuracy with minimal data, while also avoiding the need to update batch normalization (BN) parameters during training. As a result, our approach offers a low-computation, low-memory solution for efficient calibration.
\end{itemize}

\section{Background}

\subsection{Compact Model for RRAM}

In this research, we map the weight values of neural networks to the conductance values in RRAM for RIMC. In RRAM, the conductance is controlled by applying an electric field. Once the field is removed, the conductance undergoes a relaxation process, drifting over time leading to deviations between the actual conductance ($G_r$) and the target conductance ($G_t$), which negatively affects the accuracy of DNNs.


To model this drift, we assume that the deviation in conductance, $G_{drift}$, follows a Gaussian distribution \cite{Gaussian}. $G_{drift} \sim \mathcal{N}(\mu, \sigma^2)$. As a result, the actual weight values ($W_r$), derived from the conductance values, may differ from the target weight values ($W_t$). The relationship between the target and actual conductance is given by:
\begin{equation}
G_r = G_t + G_{drift}\label{eq1}
\end{equation}

The weights are linearly scaled to align with the full conductance range $G_{max}$ of the hardware. These weights are then programmed as the differential conductance $(G_r^+ - G_r^-)$ between two RRAM devices: 
\begin{equation}
W_r = (G_r^+ - G_r^-) * \frac{W_{max}}{G_{max}}\label{eq2}
\end{equation}

\begin{figure}[t]
    \centering
    \includegraphics[width=0.5\textwidth, keepaspectratio]{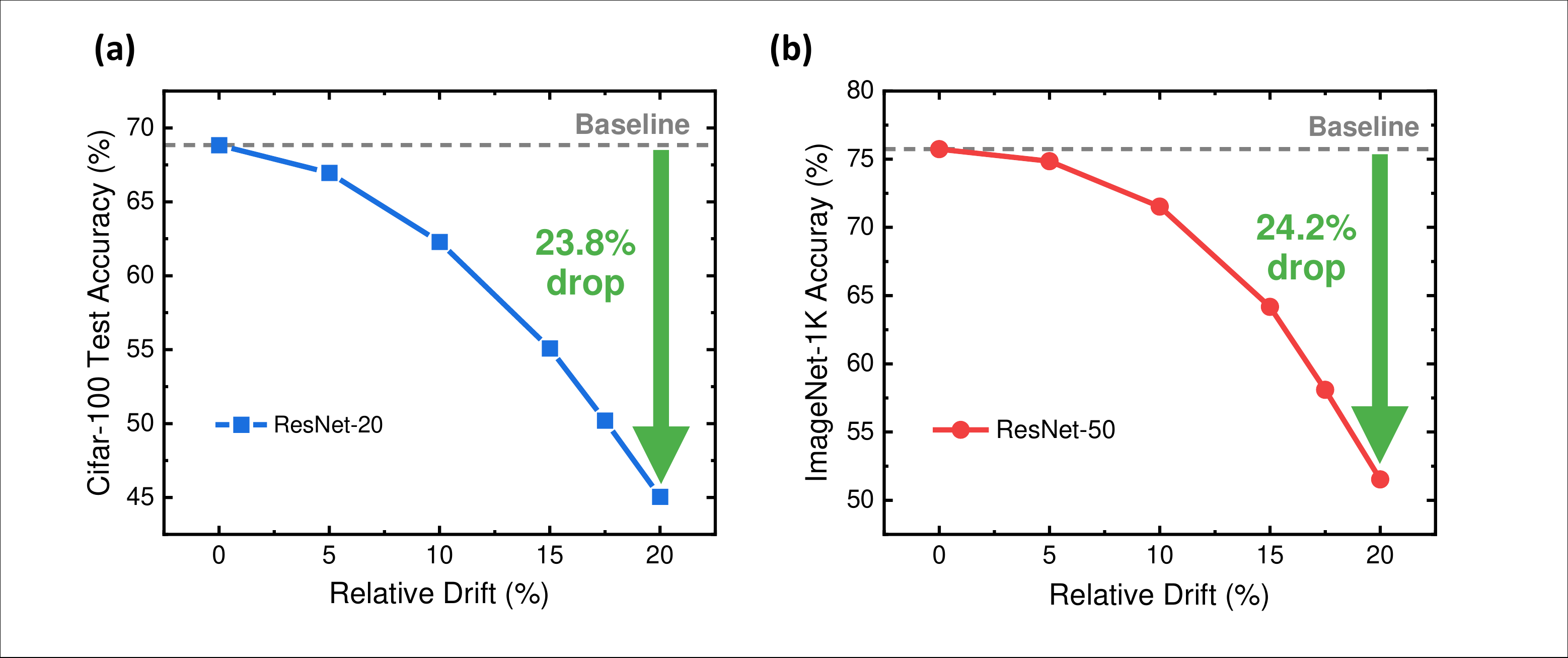}
    \caption{Impact of RRAM conductance Relaxation on (a) ResNet-20 and (b) ResNet-50. \text{Relative\ Drift} =$\frac{\sigma}{{G_{t}}}$ }
    \label{fig2}
\end{figure}

Typically, the conductance drift is large initially but stabilizes over time. However, even after stabilization, the drift still impacts the performance of RRAM-based systems. For exsiting RRAM technologies, the magnitude of drift $G_{drift}$ in conductance is generally less than of 20\% $G_t$\cite{benchmark}. As the drift increases, the performance of the DNNs degrades accordingly. When the drift is small, the impact on accuracy is relatively minor, but as the drift magnitude increase, the accuracy degradation becomes more pronounced, as illustrated in Fig.~\ref{fig2}.

Therefore, to maintain neural network accuracy in RRAM-based systems, continuous calibration is required to account for and correct these deviations over time.

\subsection{Challenges of Backpropagation-based Calibration}

To mitigate the negative impact of conductance drift on the performance of neural networks after deployment on in-memory computing chips, retraining techniques \cite{onchip}\cite{onchip2} are employed to enhance the accuracy of neural networks. However, they suffer from the following four issues.

\paragraph{Large computational cost}  
As shown in Fig.~\ref{fig2}, traditional retraining involves forward propagation to compute the network output and backpropagation to minimize the loss by updating weights. For a neural network layer \(i\), backpropagation requires calculating the gradients of the loss \(L\) with respect to the inputs (\(a_i\)) and weights (\(W_i\)), based on the forward equation \(a_{i+1} = a_i W + b\). The gradients are computed as:  
\begin{equation}
\frac{\partial L}{\partial a_i} = \frac{\partial L}{\partial a_{i+1}} W^T
\end{equation}

\begin{equation}
\quad  
\frac{\partial L}{\partial W} = a_i^T \frac{\partial L}{\partial a_{i+1}}
\end{equation}

This process demands substantial memory, as it requires storing weights, activations, optimizer states, and gradients across all layers. For high-precision formats like FP32, the need for cross-layer gradient propagation along with the updating of BN parameters, exacerbates memory and computation bottlenecks, making traditional backpropagation inefficient for resource-constrained systems. Additionally, during calibration, the BN parameters also need to be updated, further increasing memory and computational costs. This additional requirement for BN parameter updates is typically unnecessary for inference and becomes overly expensive in a calibration setting where minimal data and fewer computations are desirable.

\paragraph{Dependency on Large Datasets}
In conventional backpropagation for training neural networks, extensive datasets are required. For instance, the CIFAR-100 dataset contains 50,000 training images, while the ImageNet-1K dataset boasts an impressive 1.2 million images. Larger datasets help mitigate the risk of model overfitting, as the model learns from a broader range of data and is less likely to become overly adapted to noise and outliers in the training data. If a neural network has many layers or a high number of neurons per layer, the model's parameters can become quite numerous, giving it the capacity to learn every nuance of the training data, including noise and outliers, leading to overfitting. The larger the dataset needed, the greater the memory required for on-chip training, making it extremely challenging to perform such training on-chip.

\paragraph{Extensive Parameter Updates and RRAM Endurance Concerns}
Modern neural networks include millions to billions of parameters. For example, ResNet-20 has 268,000 parameters, ResNet-50 has 22.7 million. Traditional methods update all these weights during calibration, resulting in high computational and memory costs. Furthermore, each weight update involves programming the RRAM, which has limited write endurance. Excessive write operations degrade the resistive material, significantly reducing the lifespan of RRAM devices. This creates a trade-off between calibration frequency and device longevity, making frequent updates unfeasible.

\paragraph{Slow Training Speed}  
Due to the slow writing speed of RRAM and the need for write-and-verify operations, training times are significantly extended. The write-and-verify process, which takes approximately 100 nanoseconds per operation \cite{latency}, involves writing a new conductance value to each individual RRAM cell and verifying its correctness. If the write does not meet the target value, the process is repeated for that specific cell. As RRAM updates are performed cell by cell, this iterative nature of the process further extends training times. For instance, with ResNet-50, which has 25.6 million parameters, updating the RRAM-based ResNet-50 takes approximately 2.56 seconds per update, primarily due to this individual write process.

\begin{figure}[t]
    \centering
    \includegraphics[width=0.5\textwidth, keepaspectratio]{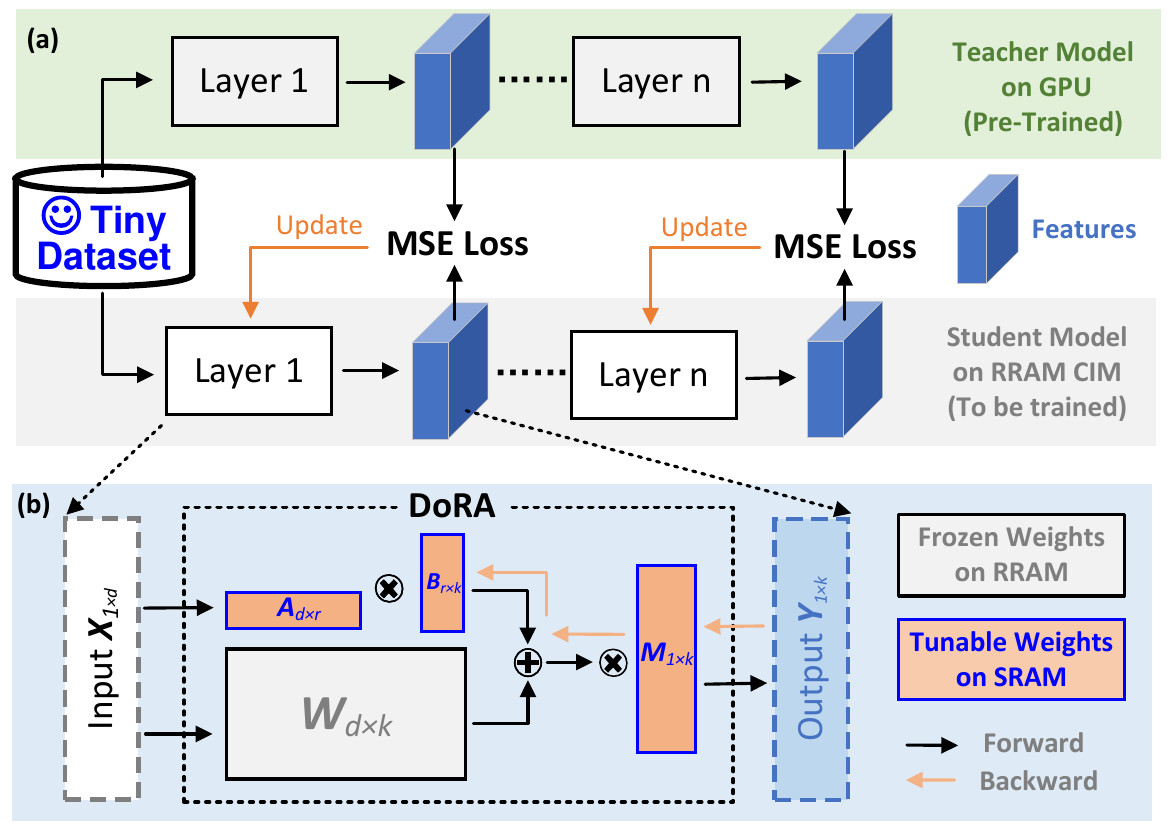}
    \caption{Overview of the proposed ultra-efficient method. (a) Feature-based calibration utilizes intermediate feature maps from a GPU-based DNN to guide the calibration of the RIMC-based DNN. This additional feature-level information reduces the risk of overfitting, enabling calibration with a tiny dataset. Calibration is performed layerwise for the RIMC. (b) Principle of DoRA: DoRA parameters are stored in digital memory, accounting for only 2.34\% of total parameters. During training, only DoRA parameters are updated, while the RRAM weights remain unchanged. }
    \label{fig3}
\end{figure}

\section{Proposed Method}
\subsection{Overall Architecture}
In this section, we propose a novel feature alignment approach that leverages feature-based KD for calibrating neural networks on RIMC hardware. As shown in Fig.~\ref{fig3}, our method is inspired by the traditional feature-based KD, where intermediate feature maps from the teacher model guide the calibration of the RIMC-based DNN. In this case, the teacher model is a DNN trained on GPU, and the student model is implemented on RIMC hardware.

The proposed method adopts a layer-wise update strategy instead of backpropagation to eliminate the memory necessary for gradient computation.
Each layer is calibrated individually with the feature-based KD.
Furthermore, since the  DoRA parameters  are stored in SRAM, updates are performed using SRAM's faster write speeds rather than the slower RRAM, which avoids the bottleneck of long programming times typically associated with RRAM write operations.

\subsection{Feature-based Calibration}\label{AA}

Feature-based calibration, as shown in Fig.~\ref{fig3}(a), is a variant of feature-based knowledge distillation where the student model, deployed on RIMC hardware, learns to align its intermediate feature representations with those of the teacher model, layer by layer. The teacher model, a DNN trained on GPU, supervises the student by guiding it to adjust its internal activations, ensuring that the feature distributions between the two models match. This method offers a key advantage by minimizing the required calibration data \cite{KD1}, which reduces computational overhead, making it particularly suitable for edge devices with limited resources.

In this approach, the student model is trained to match the feature maps of the teacher model at each layer. The training process involves minimizing the difference between the feature maps at corresponding layers of both models. Specifically, the loss function at each layer is computed as the mean squared error (MSE) between the feature map \( F_{\text{student}}^l \) of the student model and the corresponding feature map \( F_{\text{teacher}}^l \) from the teacher model. The objective is to minimize this loss to align the feature distributions, thereby transferring the essential knowledge from the teacher to the student.

The formal procedure begins with the teacher model being trained on digital hardware, denoted by the weights \( W_t \). The student model is then initialized with weights \( W_r \) and deployed on RIMC hardware, marking the start of the calibration process. During training, for each layer \( l \), the feature map \( F_{\text{student}}^l \) of the student model is computed during forward propagation, while the feature map \( F_{\text{teacher}}^l \) from the teacher is retrieved. The MSE loss between these two feature maps is calculated and used to update the parameters of the student model. The parameters of the DoRA matrices, denoted by \( A \), \( B \) and \( M \), are updated to minimize the MSE loss. These updates, which are explained in more detail later, enable efficient adaptation of the student model on the RIMC hardware.

This process is repeated for all layers in the model until convergence is achieved, defined by a threshold on the loss or a maximum number of epochs. The entire training procedure is summarized in Algorithm.~\ref{alg1}, which outlines the steps involved in the feature-based calibration process.

\begin{algorithm} 	
    \renewcommand{\algorithmicrequire}{\textbf{Input:}} 	
    \renewcommand{\algorithmicensure}{\textbf{Output:}} 	
    \caption{Feature-based Calibration } 	
    \label{alg1} 	
    \begin{algorithmic}[1] 		
        \STATE Train the DNN models on digital hardware: $W_{t}$              
        \STATE Programmed the DNN models on RMIC hardware: $W_{r}$  		
        \STATE Calculate the input and feature (output) of each layer: $F_{teacher}$              
        \FOR{each layer $l$} 		
            \REPEAT 		
                \STATE Calculate the feature of layer $l$:  $F_{student}$  		
                \STATE Calculate the loss of $l$: $\text{loss} = \text{MSE}(F_{teacher}, F_{student})$ 		
                \STATE Calculate the gradient of weight             
                \STATE Update DoRA matrices $A$, $B$, and $M$ 		 		
            \UNTIL ($\text{loss} \leq \text{threshold}$) or ($\text{epoch} \geq N$)               
        \ENDFOR 		
        \ENSURE Parameters of DoRA matrices $A$, $B$, and $M$ 	
    \end{algorithmic}   
\end{algorithm}

Feature-based calibration offers several distinct advantages over traditional calibration methods. First, by aligning feature maps rather than final predictions, the method significantly reduces the amount of calibration data required \cite{lessdata}. This reduction arises because feature maps contain richer, intermediate representations of the input data, capturing detailed, layer-wise information about the model's learned features. Aligning these intermediate representations allows the calibration process to leverage the structural consistency of the model across layers, rather than relying solely on end-to-end output predictions. As a result, even a small subset of data can effectively guide the model to correct its internal activations, reducing the need for large datasets. Additionally, because feature-based calibration occurs layer by layer, it is akin to training a neural network with only one layer at a time. This approach eliminates the need for updating BN parameters, necessary in deep network training to stabilize the learning process.
%

\subsection{Weight-Decomposed Low-Rank Adaptation}

Low Rank Adaption (LoRA) \cite{LoRA} has emerged as a powerful technique for efficiently fine-tuning large models. LoRA  modify only a small subset of model parameters, providing an efficient way to adapt pre-trained models for specific tasks without the need to retrain all parameters.

During training, the input vector \( X \in \mathbb{R}^{1 \times d} \) is passed through the original weight matrix \( W \in \mathbb{R}^{d \times k} \) as well as the newly introduced matrices \( A \in \mathbb{R}^{d \times r} \) and \( B \in \mathbb{R}^{r \times k} \), with the results combined via addition:
\begin{equation}
Y = XW + (XA)B \label{eq}
\end{equation}

In this formulation, \( A \) and \( B \) are low-rank matrices, with \( r \ll m, d \), meaning they contain far fewer parameters compared to the original weight matrix \( W \). This low-rank design minimizes the computational and memory overhead associated with training. Instead of retraining the RIMC's fixed weights stored in RRAM, we update only \( A \) and \( B \), which are stored in fast-access SRAM. This approach allows us to achieve high calibration accuracy efficiently without the need to write RRAM.

\begin{algorithm}[H]
    \renewcommand{\algorithmicensure}{\textbf{Output:}}
    \caption{DoRA}
    \label{alg2}
    \begin{algorithmic}[1]
        \STATE Programmed the DNN models on RMIC hardware: $W$
        \STATE Initialize $A \in \mathbb{R}^{d \times r}$ with random values, 
        $B \in \mathbb{R}^{r \times k}$ with zeros, and 
        $M \in \mathbb{R}^{1 \times k}$ as L2 norm of $W$ ($M = \|W\|_2$)
        \FOR{each layer $l$ in the network}
            \REPEAT
                \STATE Compute $Adapt = XW + (XA)B$
                \STATE Compute the L2 norm of Adapt: $Norm\_Adapt\ = Adapt / \|Adapt\|_2$ 
                \STATE Compute the feature of layer $l$: $F = M \circ Norm\_Adapt$
                \STATE Calculate the MSE loss $l$
                \STATE Compute the gradients for $A$, $B$, and $M$
                \STATE Update $A$, $B$, and $M$ using the gradients
            \UNTIL ($\text{loss} \leq \text{threshold}$) or ($\text{epoch} \geq N$)
            \STATE Merge $M$ and $\|Adapt\|_2$ for inference efficiency: $M = M \circ \|Adapt\|_2$
        \ENDFOR
        \ENSURE Parameters of DoRA matrices $A$, $B$, and $M$
    \end{algorithmic}
\end{algorithm}

However, LoRA approximates \( W \) through the matrices \( A \) and \( B \), and as a result, its ability to recover accuracy is not as strong as full-parameter fine-tuning. We conducted experiments on ResNet-20 using the CIFAR-100 dataset, and found that LoRA did not successfully restore accuracy (this will be discussed in more detail later). Therefore, this work does not use LoRA for calibration, but instead employs DoRA.

DoRA, a more recent technique built upon LoRA, offers an efficient approach that is closer to full-parameter fine-tuning. To overcome the limitations of LoRA, as shown in Fig.~\ref{fig3}(b), DoRA introduces an additional magnitude vector \( M \in \mathbb{R}^{1 \times k} \). As illustrate in Algorithm.~\ref{alg2}, incorporating \( M \) enables the model to adjust the magnitude of the output vector without altering its direction, while \( A \) and \( B \) can only adjust the direction of the output vector. This capability is not achievable with LoRA, which uses $A$ and $B$ for both magnitude and direction adjustments at the same time. The addition of $M$ provides greater flexibility, allowing the model to better approximate full-parameter fine-tuning\cite{DoRA}. Since the magnitude vector \( M \) contains only \( k \) parameters, it retains the computational efficiency of LoRA while providing the ability to fine-tune the output magnitude, thus significantly enhancing the model's adaptability.
\begin{equation}
Y = M \circ\left( XW + (XA)B \right) \label{eq6}
\end{equation}
Here, $\circ$ denotes the element-wise multiplication (Hadamard product). During training, only the parameters \( A \), \( B \), and \( M \) are updated. These parameters are stored in FP32 format during training, while they are quantized to integer values (int8) during inference.

\begin{table*}[t]
\centering
\caption{Performance comparison of backpropagation and our method on ImageNet-1K}
\vspace{5pt}  
\begin{tabular}{|p{2cm}|p{2cm}|p{3.4cm}|p{4cm}|p{3cm}|}
\hline
\textbf{\textit{Method}} & \textbf{\textit{Dataset Size}} & \textbf{\textit{Parameters Require Training}} & \textbf{\textit{Speed (limited by weight updates)}} & \textbf{\textit{RRAM Lifespan}} \\
\hline
Backpropagation & 125 & 100\% & Slow (1x) & 41667 calibrations\\
\hline
This Work &  10 & 2.34\%  & Fast (1250x) & $5 \times 10^{13}$\ calibrations \\
\hline
\end{tabular}
\label{tab1}
\end{table*}

Despite introducing the additional magnitude vector, DoRA maintains a significantly lower computational overhead compared to full-parameter fine-tuning. The rank \( r \) is typically much smaller than the dimensions \( d \) and \( k \) (usually $r = 4, 8$), making the number of parameters in \( A \) and \( B \) substantially smaller than in the original weight matrix \( W \), thus reducing the training burden. 
\begin{equation}
\gamma = \frac{\text{new parameters}}{\text{original parameters}} = \frac{d*r + r*k + k}{{d*k}} \label{eq7}
\end{equation}
%
where \( \gamma \) represents the proportion of new parameters introduced by DoRA training relative to the original network, highlighting its efficiency.
\begin{figure}[t]
    \centering
    \includegraphics[width=0.5\textwidth, keepaspectratio]{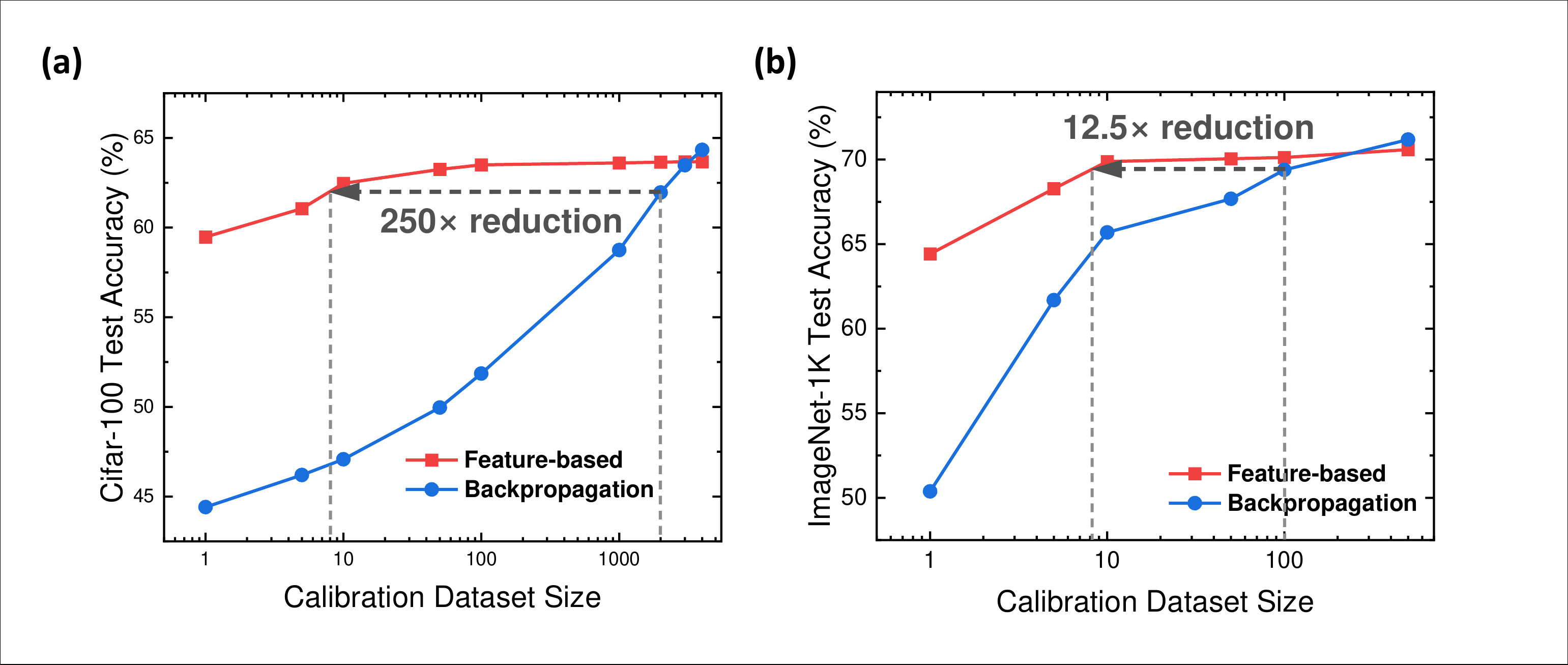}
    \caption{Comparison of calibration dataset size for feature-based calibration and backpropagation when relative drift is 20\% on (a) Cifar-100 dataset with r =2 and (b)  ImageNet-1K dataset with r = 4.  }
    \label{fig4}
\end{figure}

\section{Experimental Results}
\subsection{Experimental Setup}
We evaluated both conventional backpropagation calibration, and feature-based KD enhanced with DoRA—on two datasets: CIFAR-100 using ResNet-20 and ImageNet-1K using ResNet-50. The evaluation metrics included accuracy, memory usage, required dataset size, and the number of trainable parameters.

For baseline accuracy, the ResNet-20 model achieved 65.6\% top-1 accuracy on CIFAR-100, and the ResNet-50 model achieved 75.73\% top-1 accuracy on ImageNet-1K. These models served as the pre-trained networks for our experiments. Subsequently, we extracted various parameters, including weights and intermediate feature maps, from these trained networks. To simulate the impact of RRAM conductance drift, we perturbed the weights using \eqref{eq2} to introduce Gaussian noise corresponding to typical drift levels. 

For backpropagation-based method, training was conducted using the final output of the network as the guidance signal, with cross-entropy loss serving as the objective function. In contrast, for the feature-based methods, intermediate feature maps were used to calculate the loss using the MSE metric, allowing for more granular guidance during training. All models were fine-tuned for 20 epochs on a subset of the respective datasets to simulate real-world constraints on calibration data availability.

\begin{figure}[t]
    \centering
    \includegraphics[width=0.5\textwidth, keepaspectratio]{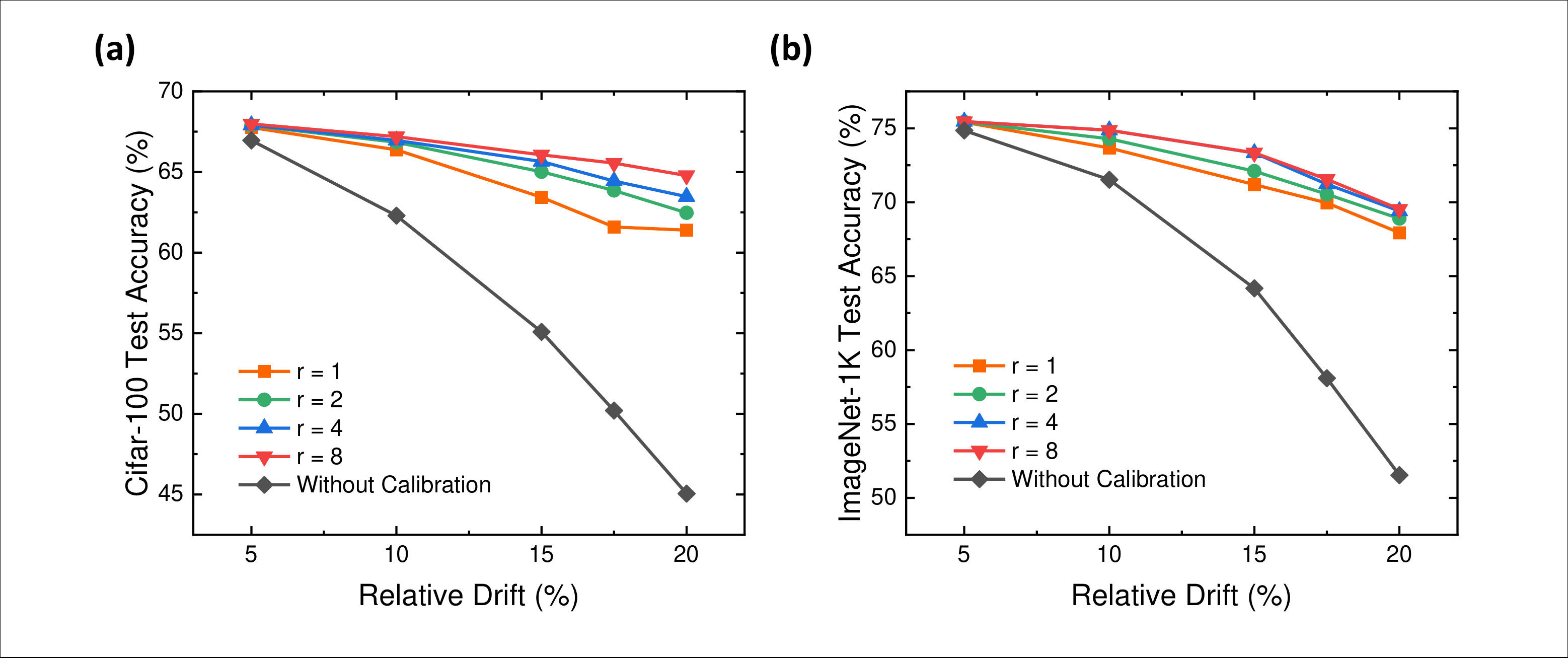}
    \caption{Comparison of calibration dataset size for feature-based calibration and backpropagation when relative drift is 20\% on (a) Cifar-100 dataset and (b)  ImageNet-1K dataset.  }
    \label{fig5}
\end{figure}

\subsection{Size of Calibration Dataset}
As shown in Fig.~\ref{fig4}, even when using an extremely small calibration dataset, we observed that feature-based calibration did not exhibit significant overfitting. Specifically, with just a single calibration sample, ResNet20 achieved an impressive accuracy of 58.44\%, which is a substantial improvement over the pre-calibration accuracy of 45.05\%. In contrast, backpropagation with a single calibration sample resulted in only 44.01\% accuracy, even lower than the pre-calibration accuracy. This clearly demonstrates the effectiveness of our method, even with minimal data. Our method, with a calibration dataset size of 10 samples (achieving 63.55\% accuracy on CIFAR-100 and 69.53\% on ImageNet-1k), clearly outperforms the backpropagation method with the same dataset size (achieving 47.10\% accuracy on CIFAR-100 and 65.79\% on ImageNet). The performance of our method is very close to that of backpropagation using 2000 samples on CIFAR-100 and 100 samples on ImageNet-1k, reinforcing the effectiveness of our approach in scenarios with small calibration datasets.

Despite the large improvement seen with a small number of calibration samples, using a large dataset for calibration is impractical in RIMC contexts. Feature-based calibration thus remains a clear advantage, offering a practical solution for scenarios where dataset size is limited.

\subsection{Impact of Rank $r$}

The relationship between the size of parameter $r$ and the post-calibration accuracy is delineated in Fig.~\ref{fig5}. A larger $r$ can yield improved post-calibration accuracy. However, according to \eqref{eq7}, the parameter overhead increases linearly with the increase in $r$. Consequently, there is a trade-off to consider when performing calibration. Small $r$ is ideal for lightweight deployments with minor RRAM conductance drift, achieving faster inference and low overhead. Large $r$ is suitable for scenarios with significant RRAM conductance drift, though it increases parameter and computation costs. By balancing $r$, DoRA enables precise, efficient corrections tailored to specific application constraints, demonstrating its utility in RIMC.

For instance, as the model size increases (i.e., the values of \( d \) and \( k \) become larger in ResNet-50), the size of the weight matrix \( W \in \mathbb{R}^{d \times k} \) grows significantly, making the product \( d \times k \) much larger than \( d + k \). According to Eq. (9), the proportion of new parameters added through the low-rank approximation is smaller in larger models. This is because the new parameters introduced by matrices \( A \) and \( B \) are only a fraction of the original matrix, and as the model grows in size, this fraction decreases.

For example, when \( r = 1 \), we find that the added parameter proportion in ResNet-20 is 4.46\%, while in ResNet-50, it is only 0.585\%. This shows that the overhead (in terms of the proportion of new parameters) becomes much smaller in larger models, making DoRA more computationally efficient relative to the model size.

\subsection{Lifespan} 
Given that the endurance of RRAM is \(10^8\) cycles, and during a single backpropagation-based calibration there are 20 epochs and 120 calibration samples, with a batch size of 1 to simulate resource-constrained conditions, the RRAM conductance values are updated 120 times per epoch. Thus, in one calibration, the RRAM is updated 2400 times. This means the RRAM can undergo a total of 41,667 calibrations.

In contrast, the DoRA-based method does not require updating the RRAM, instead updating the SRAM, which typically has an endurance of \(10^{16}\) cycles. With 10 data samples, each calibration update involves 200 SRAM updates, allowing for a total of \(5 \times 10^{13}\) calibrations.

\subsection{Training Speed Result}

The training speed difference between backpropagation and our method is mainly determined by the weight update time, as the computation time in both approaches is similar.

In our method, we achieve comparable calibration accuracy with only 8\% of the original calibration dataset. Consequently, for a given number of epochs and batch size, the number of updates required is only 8\% of that needed by traditional backpropagation.

Furthermore, since RRAM write time is approximately 100 times slower than SRAM, the time spent on weight updates in our method (using SRAM) is only 0.08\% of the time required for backpropagation using RRAM, making our method 1250 times faster than traditional backpropagation method.

\begin{figure}[t]
    \centering
    \includegraphics[width=0.5\textwidth, keepaspectratio]{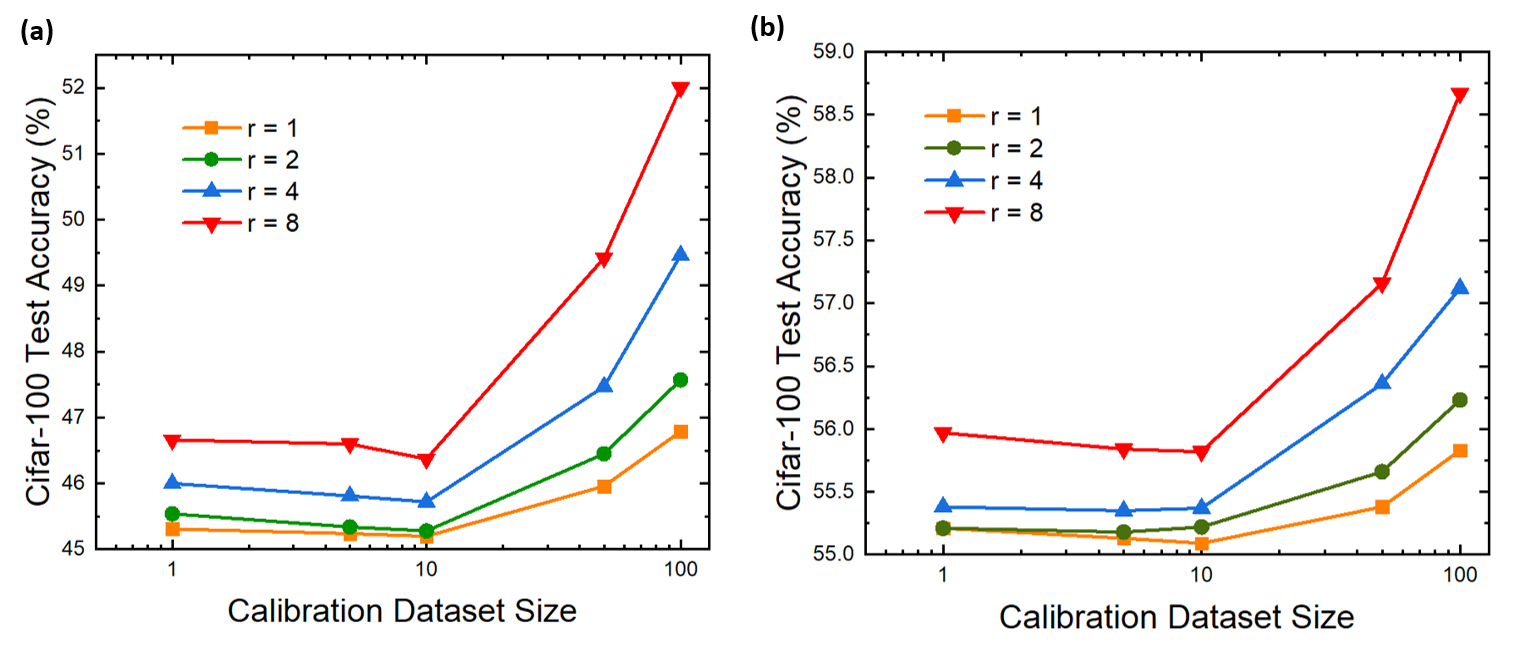}
    \caption{The result of LoRA enhanced feature-based calibration when on Cifar-100 dataset when relative drift is (a) 20\% (b) 15\%.  }
    \label{fig6}
\end{figure}

\subsection{Comparison with LoRA} 
As shown in Fig.~\ref{fig6}, we tested the impact of using LoRA instead of DoRA on post-calibration accuracy. We found that LoRA has relatively limited calibration adjustment capability. When the relative drift is set to 0.2, the lowest accuracy after DoRA calibration, achieved with a rank of 1, is 61.39\%, which outperforms the highest accuracy of 52.11\% obtained with a rank of 8 in LoRA. The same trend is observed when the relative drift is 0.15. In contrast, DoRA proves to be a more suitable calibration technique.

\subsection{Comparison with Backpropagation}

In Table \ref{tab1}, a detailed comparison of the performance between backpropagation and our method is presented. Our approach shows significant advantages in terms of dataset size, the proportion of tunable parameters, calibration speed, and RRAM friendliness. One limitation of our method is that we introduce the DoRA, which introduces some overhead.

\section{Conclusion}
This work presents an efficient calibration framework designed to enhance the reliability and longevity of RIMC systems. By addressing key challenges such as conductance relaxation, high write time consumption, and limited endurance, our method ensures sustained DNNs performance with minimal calibration costs. Through the integration of a feature-based calibration strategy and DoRA, we eliminate the need for repeated RRAM writes by shifting tunable parameter storage and calibration computation to digital memory. Experimental results on an RIMC-based ResNet-50 using ImageNet-1K confirm the effectiveness of our approach, achieving 69.53\% accuracy restoration with just 10 calibration samples and 2.34\% parameters need to be trained, significantly reducing the computational and energy overhead. Our framework demonstrates the potential for scalable and resource-efficient calibration solutions critical for managing the silicon lifecycle of RIMC-based systems, paving the way for their broader adoption in edge AI and IoT applications.

\end{document}